\documentclass[%
reprint,
superscriptaddress,
amsmath,amssymb,
prb,
longbibliography
]{revtex4-2}

\usepackage{color}
\usepackage{xcolor}
\usepackage{graphicx}
\usepackage{dcolumn}
\usepackage{siunitx}
\usepackage{bm}
\usepackage{hyperref}
\hypersetup{colorlinks=true,linkcolor=blue,citecolor=blue,urlcolor=blue}
\usepackage{soul} 
\usepackage[colorinlistoftodos]{todonotes}
\usepackage{mhchem}

\begin{document}

\title{Single-crystal growth, structural characterization, and physical properties of a decorated square-kagome antiferromagnet
KCu$_7$TeO$_4$(SO$_4$)$_5$Cl}

\author{Jingjing Jing}
\thanks{These authors contributed equally to this work. }
\affiliation{Low Temperature Physics Laboratory, College of Physics \& Center of Quantum Materials and Devices, Chongqing University, Chongqing 401331, China}

\author{Andreas Eich}
\thanks{These authors contributed equally to this work. }
\affiliation{Institute for Quantum Materials and Technologies, Karlsruhe Institute of Technology, Kaiserstraße 12, 76131 Karlsruhe, Germany}

\author{Yiqiu Liu}
\thanks{These authors contributed equally to this work. }
\affiliation{School of Mathematics and Physics,~North China Electric Power University,~Beijing,~102206,~China}
\affiliation{School of Physics and Beijing Key Laboratory of Opto-Electronic Functional Materials \& Micro-Nano Devices, Renmin University of China, Beijing, 100872, China}


\author{Liran Wang}
\affiliation{Institute for Quantum Materials and Technologies, Karlsruhe Institute of Technology, Kaiserstraße 12, 76131 Karlsruhe, Germany}


\author{Lunhua He}
\affiliation{Beijing National Laboratory for Condensed Matter Physics, Institute of Physics, Chinese Academy of Sciences, Beijing 100190, China}
\affiliation{Spallation Neutron Source Science Center, Dongguan, 523803,China}

\author{Aifeng Wang}
\affiliation{Low Temperature Physics Laboratory, College of Physics \& Center of Quantum Materials and Devices, Chongqing University, Chongqing 401331, China}

\author{Yisheng Chai}
\affiliation{Low Temperature Physics Laboratory, College of Physics \& Center of Quantum Materials and Devices, Chongqing University, Chongqing 401331, China}

\author{Young Sun}
\affiliation{Low Temperature Physics Laboratory, College of Physics \& Center of Quantum Materials and Devices, Chongqing University, Chongqing 401331, China}

\author{Yi Cui}
\affiliation{School of Physics and Beijing Key Laboratory of Opto-Electronic Functional Materials \& Micro-Nano Devices, Renmin University of China, Beijing, 100872, China}

\author{Frederic Hardy}
\affiliation{Institute for Quantum Materials and Technologies, Karlsruhe Institute of Technology, Kaiserstraße 12, 76131 Karlsruhe, Germany}

\author{Christoph Meingast}
\affiliation{Institute for Quantum Materials and Technologies, Karlsruhe Institute of Technology, Kaiserstraße 12, 76131 Karlsruhe, Germany}

\author{Weiqiang Yu}
\email{wqyu\_phy@ruc.edu.cn}
\affiliation{School of Physics and Beijing Key Laboratory of Opto-Electronic Functional Materials \& Micro-Nano Devices, Renmin University of China, Beijing, 100872, China}

\author{Xinrun Mi}
\email{xinrunmi@cqu.edu.cn}
\affiliation{Low Temperature Physics Laboratory, College of Physics \& Center of Quantum Materials and Devices, Chongqing University, Chongqing 401331, China}
\affiliation{Chongqing Police College, Chongqing 401331, China}

\author{Michael Merz}
\email{michael.merz@kit.edu}
\affiliation{Institute for Quantum Materials and Technologies, Karlsruhe Institute of Technology, Kaiserstraße 12, 76131 Karlsruhe, Germany}
\affiliation{Karlsruhe Nano Micro Facility, Karlsruhe Institute of Technology, Kaiserstraße 12, 76131 Karlsruhe, Germany}

\author{Mingquan He}
\email{mingquan.he@cqu.edu.cn}
\affiliation{Low Temperature Physics Laboratory, College of Physics \& Center of Quantum Materials and Devices, Chongqing University, Chongqing 401331, China}

\date{\today}

\begin{abstract}
The square-kagome lattice antiferromagnet, composed of two-dimensional corner-sharing triangles, provides a prominent platform for studying frustrated magnetism. However, material realizations of the square-kagome lattice remain scarce. Here, we report the single-crystal growth, structural characterization, magnetic and electric properties of KCu$_7$TeO$_4$(SO$_4$)$_5$Cl, a nabokoite compound featuring a distorted and decorated square-kagome lattice. Weak anomalies at temperature near 4~K are observed in both the magnetization and the specific heat data, indicating the onset of a magnetic transition. The observation of the $^{35}$Cl NMR line splits below 4.5~K further confirms the formation of the long-range-order antiferromagnetism.  Magnetic susceptibility data reveal nearly isotropic Curie–Weiss temperatures ($\sim-145$ K) and $g$-factors ($\sim2.4$) for both in-plane and out-of-plane magnetic fields. Moreover, we observe two successive ferroelectric transitions at $T_\mathrm{FE1}\sim30$ K and $T_\mathrm{FE2}\sim27$ K, driven by inversion-symmetry breaking, most likely associated with distortions in the Cu2O$_4$Cl$_1$ pyramids and the adjacent SO$_4$ tetrahedra.  These results suggest that a three-dimensional model incorporating interlayer couplings via decorating sites is essential for capturing the magnetic and electric behaviors in KCu$_7$TeO$_4$(SO$_4$)$_5$Cl.
\end{abstract}

\maketitle

\section{introduction}

Frustrated magnetic systems have long been a central topic in condensed matter physics due to their ability to host exotic ground states that deviate from conventional long-range magnetic order \cite{anderson1987resonating,balents2010spin,ramirez1994strongly}. In these systems, competing interactions inhibit conventional symmetry breaking, giving rise to highly entangled quantum states such as quantum spin liquids \cite{savary2016quantum,zhou2017quantum}. The geometrical frustration inherent to corner-sharing triangular lattices, particularly in two-dimensional (2D) networks, serves as a fundamental mechanism in realizing such phenomena. Among the various frustrated geometries, the kagome lattice—composed of corner-sharing triangles in two dimensions—has emerged as a paradigmatic platform for studying the interplay between geometric frustration and quantum fluctuations \cite{syozi1951statistics,kano1953antiferromagnetism,depenbrock2012nature,liao2017gapless}.

\begin{figure*}[t]

\includegraphics[scale= 0.565]{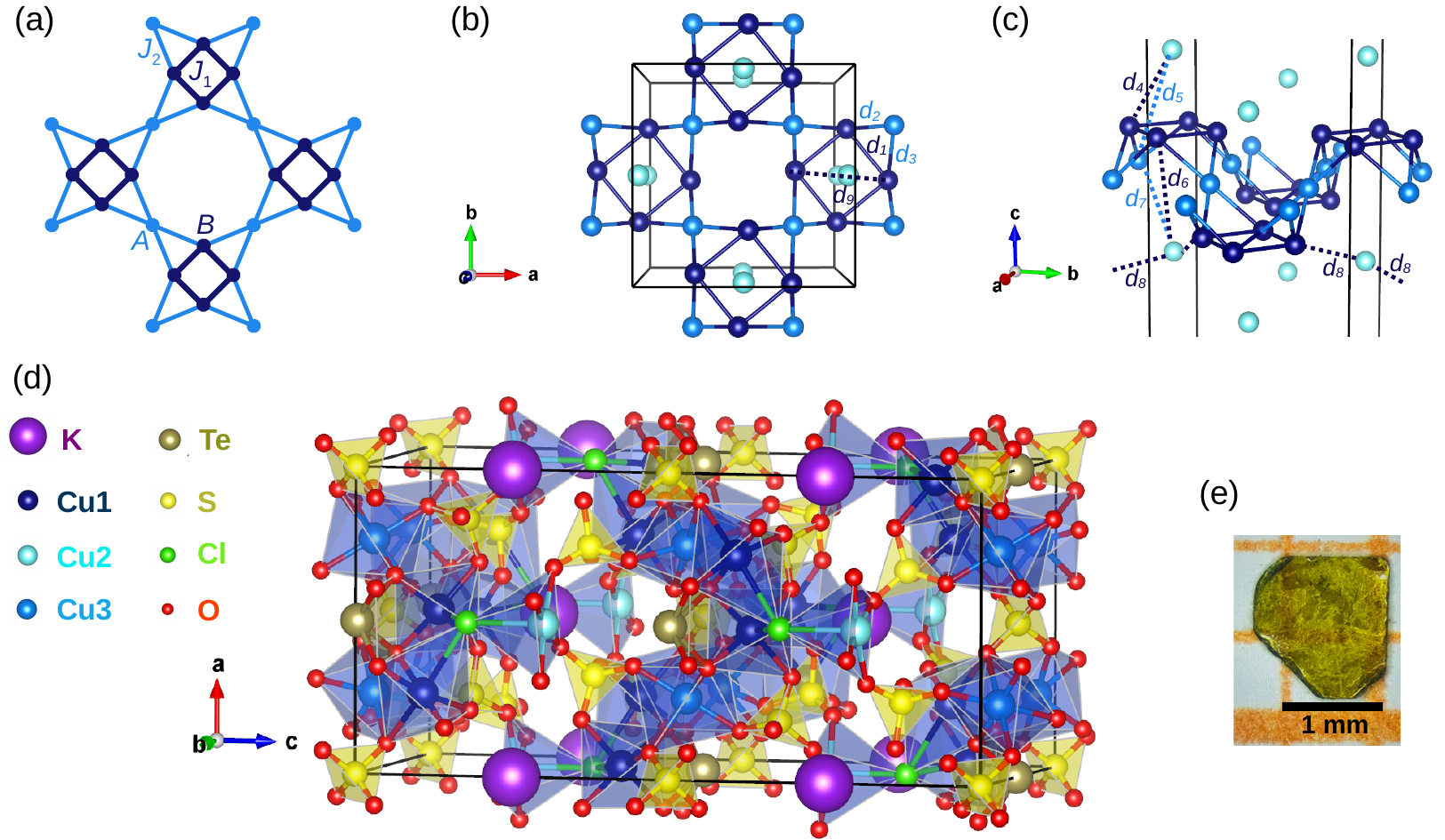}
\caption{The square kagome lattice in KCu$_7$TeO$_4$(SO$_4$)$_5$Cl: (a) Idealized theoretical picture of the SKL lattice. (b) Perspective top-down view of the SKL plaquettes featuring Cu1 square and Cu1-Cu3-Cu1 triangular motifs, highlighting the coordination between copper atoms in the lattice. Bond distances $d_1$, $d_2$, and $d_3$ as well as the diagonal distance $d_9$ are outlined. The Cu2 atoms are positioned above and below the distorted SKL plaquettes.
(c) Perspective side view, illustrating how the SKL Cu network is corrugated, with the Cu1-Cu3-Cu1 triangles alternating between pointing upwards and downwards, and the Cu2 atoms positioned above and below the Cu1 squares. Bond distances $d_4$, $d_5$, $d_6$, and $d_7$ are outlined, and the distance $d_8$ of Cu2 to the Cu1 atom situated in the neighboring corrugated SKL layer (not shown) is indicated schematically. (d) The complete KCu$_7$TeO$_4$(SO$_4$)$_5$Cl unit cell, including all present atoms and the most prominent coordination polyhedra. For clarity, the atomic positions in this figure have been shifted by $-\frac{1}{4}$, $\frac{1}{4}$, 0 relative to the refined ones in Tables \ref{Table_XRD} and \ref{Table_XRD2}.\@ (e) Representative pale green, millimeter-sized single crystals. Structural properties are visualized using \textsc{VESTA} \cite{Mommadb5098}. }  
\label{fig1}
\end{figure*}

A structural variant of the kagome lattice, known as the square-kagome lattice (SKL), was theoretically proposed by Siddharthan \textit{et al.} in 2001 as an alternative platform for exploring frustrated magnetism \cite{siddharthan2001square}. As illustrated in Fig. \ref{fig1}(a), the SKL is a two-dimensional lattice composed of corner-sharing triangles arranged with alternating square and octagonal voids \cite{siddharthan2001square}. Like the kagome lattice, the SKL features a low coordination number ($z = 4$). However, it differs in hosting two inequivalent sites, labeled A and B in Fig. \ref{fig1}(a). For a spin-1/2 system with Heisenberg antiferromagnetic (AFM) interactions on the SKL, a range of nontrivial ground states—including valence bond crystals (VBCs), quantum spin liquids (QSLs), and topological nematic spin-liquid phases—can emerge depending on the ratio of nearest-neighbor interactions [$J_1$ and $J_2$ in  Fig. \ref{fig1}(a)] \cite{Tomczak_2003,richter2004magnetic,derzhko2013frustrated,rousochatzakis2013frustrated,nakano2014spin,ralko2015resonating,pohle2016reentrance,morita2018magnetic,hasegawa2018metamagnetic,Astrakhantsev_singlet,richter2022thermodynamics}.

The experimental realization of an ideal square-kagome lattice (SKL) remains challenging. Nevertheless, several natural minerals—such as nabokoite KCu$_7$TeO$_4$(SO$_4$)$_5$Cl \cite{pertlik1988crystal}, atlasovite KCu$_6$FeBiO$_4$(SO$_4$)$_5$Cl \cite{popova1987nabokoite}, elasmochloite Na$_3$Cu$_6$BiO$_4$(SO$_4$)$_5$ \cite{pekov2019elasmochloite}, and favreauite PbCu$_6$BiO$_4$(SeO$_3$)$_4$(OH)·H$_2$O \cite{mills2014favreauite}—exhibit SKL-like structural motifs. By replacing Fe with Al in atlasovite, Fujihala \textit{et al.} successfully synthesized KCu$_6$AlBiO$_4$(SO$_4$)$_5$Cl, in which the Cu atoms form a square-kagome network \cite{fujihala2020gapless}. Interestingly, signatures of a gapless quantum spin liquid (QSL) have been reported in this compound \cite{fujihala2020gapless,liu2022low}, generating renewed interest in square-kagome systems. Using a hydrothermal method, Yakubovich \textit{et al.} synthesized Na$_6$Cu$_7$BiO$_4$(PO$_4$)$_4$[Cl,(OH)]$_3$, which hosts a square-kagome Cu network decorated by interlayer Cu atoms \cite{yakubovich2021hydrothermal}. No magnetic order has been detected above 50 mK \cite{liu2022low}, and theoretical studies suggest that interlayer Cu decoration plays a key role in stabilizing the quantum paramagnetic state in Na$_6$Cu$_7$BiO$_4$(PO$_4$)$_4$Cl$_3$ \cite{niggemann2023quantum}. More recently, Markina \textit{et al.} and Murtazoev \textit{et al.} synthesized a family of nabokoite-like compounds, $A$Cu$_7$TeO$_4$(SO$_4$)$_5$Cl ($A =$ Na, K, Rb, Cs), which also feature distorted SKL networks of Cu atoms coupled to interlayer Cu sites \cite{markina2024static,murtazoev2023new}. While magnetic order emerges below 5 K in compounds with $A=$ Na, K, and Rb, CsCu$_7$TeO$_4$(SO$_4$)$_5$Cl appears to remain non-magnetic down to 2 K \cite{murtazoev2023new}. Furthermore, theoretical work suggests the possibility of a field-induced spin liquid phase in KCu$_7$TeO$_4$(SO$_4$)$_5$Cl, highlighting this system as another promising SKL platform for investigating frustrated quantum magnetism \cite{Gonzalez2025}.

Despite the presence of interlayer decorating sites in Na$_6$Cu$_7$BiO$_4$(PO$_4$)$_4$Cl$_3$ and $A$Cu$_7$TeO$_4$(SO$_4$)$_5$Cl, earlier theoretical studies suggest that these systems are essentially quasi-two-dimensional with dominant intralayer magnetic interactions \cite{niggemann2023quantum,Gonzalez2025}. However, previous experimental studies have primarily focused on polycrystalline samples \cite{yakubovich2021hydrothermal,liu2022low,markina2024static,murtazoev2023new}, leaving the anisotropic properties of these quasi-2D materials largely unexplored. Moreover, magnetic states in frustrated systems are highly sensitive to structural details, and distortions induced by external factors—such as those from compressed powder samples used in specific heat measurements—may obscure the intrinsic behaviors of square-kagome materials.

In this work, we report a successful single-crystal growth of KCu$_7$TeO$_4$(SO$_4$)$_5$Cl, the nabokoite compound featuring distorted square-kagome lattice. Single crystals were synthesized via the chemical vapor transport (CVT) technique, and the crystal structure was determined by single-crystal x-ray diffraction, confirming a layered network of Cu$^{2+}$ ions forming a distorted SKL framework decorated by interlayer Cu sites. To probe the magnetic properties, we performed magnetic susceptibility, specific heat, and nuclear magnetic resonance (NMR) measurements. The results reveal a long-range antiferromagnetic order below about 4.5 K. The anisotropy is found to be weak in magnetic susceptibility with magnetic fields applied in-plane and out-of-plane, suggesting the importance of interlayer coupling in determining the magnetic properties of KCu$_7$TeO$_4$(SO$_4$)$_5$Cl. 


\section{Methods}

{\it Sample growth}. Polycrystalline samples of KCu$_7$TeO$_4$(SO$_4$)$_5$Cl were synthesized via a solid-state reaction method \cite{markina2024static}. The starting materials, in the molar ratio of anhydrous CuSO$_4$, TeO$_2$, CuO, and KCl, were used. Anhydrous CuSO$_4$ and TeO$_2$ were dried at 200 $^\circ$C for 24 hours prior to use. After thoroughly mixing the powders, the mixture was sealed in a quartz ampoule. The reaction was carried out under a dry argon atmosphere. The sample was heated at a rate of 1 $^\circ$C/min to 550 $^\circ$C and held at that temperature for one week. The sintering process was repeated multiple times to ensure the formation of high-quality polycrystalline material.

Single crystals of KCu$_7$TeO$_4$(SO$_4$)$_5$Cl were grown using a chemical vapor transport (CVT) method. The polycrystalline sample was mixed with TeCl$_4$ as a transport agent in a 10:1 weight ratio and sealed in a quartz ampoule under a dry argon atmosphere. The ampoule was then heated in a two-zone furnace, with the temperature set to 500 $^\circ$C in the low-temperature zone and 550 $^\circ$C in the high-temperature zone, and maintained for 14 days. Pale green, millimeter-sized, square-shaped single crystals were obtained [see Fig. \ref{fig1}(e)].

{\it Single crystal X-ray diffraction.} The crystal structure is probed by single-crystal x-ray diffraction (SC-XRD).\@ Measurements were performed at 300, 275, 250, 225, 200, 175, 150, 125, 100, and 80 K on a high-flux, high-resolution, rotating anode Rigaku Synergy-DW (Mo/Ag) diffractometer using Mo $K_\mathrm{\alpha}$ radiation ($\lambda$ = 0.7107 {\AA}). The system is equipped with pairs of precisely manufactured Montel mirror optics, a motorized divergence slit which was set to 5 mrad, and a background-less Hypix-Arc150$^{\circ}$ detector which guarantees the lowest reflection profile distortion and ensures that all reflections are detected under equivalent conditions.
The specimen had a size of $\approx$ 30 $\times$ 30 $\times$ 20 \textmu m$^3$ and was investigated to a resolution better than 0.5 {\AA} and exhibited no mosaic spread and no additional reflections from secondary phases, highlighting its high quality and allowing for excellent evaluation using the latest version of the \textsc{CrysAlisPro} software package \cite{CrysAlis}. The crystal structure was initially solved using the charge-flipping algorithm implemented in \textsc{Superflip}~\cite{palatinus2007superflip} and subsequently refined using \textsc{JANA2020} \cite{PetricekPalatinusPlasilDusek+2023+271+282},  including all averaged symmetry-independent reflections (I $>$ 2 $\sigma$). Unit cell and space group were determined, atoms were localized within the unit cell using random phases and the structure was completed and finally solved using difference Fourier analysis. The structural refinements converged well, exhibiting excellent reliability factors (see Tables \ref{Table_XRD} and \ref{Table_XRD2} for residuals $wR_\mathrm{2}$,  $R_\mathrm{1}$,  and goodness of fit, GOF,  values).

{\it Magnetization and specific heat.} Magnetization measurements were performed using a Quantum Design Magnetic Property Measurement System (MPMS), while specific heat measurements were carried out using a Physical Property Measurement System (PPMS - 9 T).

{\it Dielectric constant and pyroelectric current measurements.} The complex dielectric constant ($\varepsilon=\varepsilon'+i\varepsilon''$, where $\varepsilon'$ and $\varepsilon''$ are real and imaginary parts of the complex dielectric constant) of KCu$_7$TeO$_4$(SO$_4$)$_5$Cl single crystals were measured using an Agilent E4980A LCR meter. The electrodes were prepared by gluing silver expoxy to the surfaces perpendicular to the $c-$axis with electric fields applied parallel to the $c-$axis. The real part of the relative dielectric constant is evaluated via $\varepsilon'=C_\mathrm{e}d/\varepsilon_0A$, where $\varepsilon_0$, $C_\mathrm{e}$, $d$, $A$ represent the dielectric constant of vacuum, sample electric capacitance, thickness and area of electrodes, respectively. The pyroelectric current of KCu$_7$TeO$_4$(SO$_4$)$_5$Cl was measured using a Keithley 6517B electrometer. The background current was initially measured during heating (3 K/min) after cooling the samples down to 2 K in zero applied electric field. A poling electric field of 10 kV/cm was then applied at 60 K using the Keithley 6517B, and the samples were cooled to 2 K under this electric field. After removing the electric field, the electrodes were short-circuited for 30 minutes at 2 K to allow the release of any trapped charge. The samples were subsequently heated at a rate of 3 K/min, and the pyroelectric current was recorded during this process. The electric polarization $P=1/A\int I_\mathrm{p}(t)dt$ was calculated by integrating the time-dependent pyroelectric current $I_\mathrm{p}(t)$.

{\it Polarization–electric field (P–E) loop measurements.} $P–E$ loops were measured using a TF Analyzer 2000E in the static hysteresis configuration. A triangular electric-field waveform at 10 Hz was applied, and the polarization $P$ was recorded as a function of the applied electric field $E$ to characterize the ferroelectric response.

{\it Nuclear magnetic resonance.} 
NMR measurements were performed on $^{35}$Cl nuclei ($I=3/2$ and Zeeman factor $^{35}\gamma$~=~4.171~MHz/T). The NMR spectra were collected by the standard spin-echo technique. The spin-lattice relaxation rate $1/T_1$ was measured by the inversion-recovery method, with the spin recovery curve fit to the exponential function for spin-3/2 nuclei, $I(t)/I(\infty)=1-a[0.1e^{(-t/T_1)^\beta}+0.9e^{-(6t/T_1)^\beta}]$ ($1<a<2$), where $\beta$ is a stretching factor. $\beta \approx 1$ in the paramagnetic phase, indicating high-quality of the sample.

{\it Thermal expansion.} Thermal-expansion experiments were carried out using a homebuilt high-resolution capacitive dilatometer \cite{Meingast1990}.

\begin{table*}[t]
\caption{Temperature-dependent crystallographic data for KCu$_7$TeO$_4$(SO$_4$)$_5$Cl determined from single-crystal x-ray diffraction. The structure was refined in the tetragonal space group (SG) $P4/ncc$ for which the lattice parameters $a$ and $c$  are shown together with the volume $V$ of the unit cell. The coordinates of all atoms in KCu$_7$TeO$_4$(SO$_4$)$_5$Cl are given together with their corresponding Wyckoff positions. U$_{eq}$ denotes the equivalent atomic displacement parameters (ADP). The ADPs were refined anisotropically but due to space limitations only the $U_{\rm eq}$ are listed in the Table. Likewise, the refinements of the data measured at 275 K and 225 K were omitted. Errors shown are statistical errors from the refinement. 
}
\begin{ruledtabular}
		\begin{tabular}[t]{lccccccc}
                $T$ (K)     &             300          &          250       &       200     &   150  &    125  &    100  &    80  \\ \hline
                $a$ (\AA)              &    9.7972(1)       &     9.7933(1)    &       9.7884(1)  & 9.7844(1) & 9.7825(1) & 9.7807(1) & 9.7795(1) \\
			  $c$ (\AA)      &        20.5189(5)     &       20.5045(2)   &       20.4890(1)  & 20.4750(1) & 20.4682(1) & 20.4628(1) & 20.4604(1) \\
			  $V$ (\AA$^3$)            &      1969.5           &     1966.5       &      1963.1 & 1960.2 & 1958.7 & 1957.5 & 1956.8 \\ 
       \multicolumn{8}{c}{Cu1, $16g$: $x, y, z$ }  \\
       	 $x$                 &    0.48509(1)        &      0.48508(2)    &      0.48502(1)  & 0.48497(1) & 0.48495(1) & 0.48492(1) & 0.48491(1) \\
       	 $y$                 &    0.27427(2)        &      0.27443(2)    &      0.27464(1)  & 0.27482(1) & 0.27490(1) & 0.27500(1) & 0.27505(1) \\
      	 $z$                 &    0.89481(1)        &      0.89476(1)    &      0.89477(1)  & 0.89475(1) & 0.89473(1) & 0.89472(1) & 0.89471(1) \\
				 $U_{\rm eq}$ (\AA$^2$) &   0.00922(3)     &      0.00782(4)   &     0.00667(3) & 0.00538(2) & 0.00474(2) & 0.00416(3) & 0.00347(3) \\ 
    \multicolumn{8}{c}{Cu2, $4c$: $\frac{1}{4}, \frac{1}{4}, z$}  \\
      	 $z$                 &    0.70001(1)        &      0.70005(2)    &      0.70020(1)  & 0.70031(1) & 0.70036(1) & 0.70040(1) & 0.70044(1) \\
				 $U_{\rm eq}$ (\AA$^2$) &   0.00831(4)     &      0.00719(6)   &     0.00603(3) & 0.00486(3) & 0.00430(3) & 0.00379(3) & 0.00315(3) \\ 
    \multicolumn{8}{c}{Cu3, $8d$: $0, 0, 0$}  \\
				 $U_{\rm eq}$ (\AA$^2$) &   0.01011(5)     &      0.00863(6)   &     0.00729(4) & 0.00586(4) & 0.00517(4) & 0.00452(4) & 0.00377(4) \\ 
    \multicolumn{8}{c}{Te, $4c$: $\frac{1}{4}, \frac{1}{4}, z$}  \\
      	 $z$                 &    0.99713(1)        &       0.99716(2)    &      0.99722(1)  & 0.99726(1) & 0.99729(1) & 0.99732(1) & 0.99733(1) \\
				 $U_{\rm eq}$ (\AA$^2$) &   0.00633(4)     &      0.00544(3)   &     0.00473(2) & 0.00387(2) & 0.00343(2) & 0.00303(2) & 0.00247(2) \\ 
       \multicolumn{8}{c}{S1, $16g$: $x, y, z$ }  \\
       	 $x$                 &    0.56899(3)        &      0.56916(4)    &      0.56926(2)  & 0.56937(2) & 0.56942(2) & 0.56946(2) & 0.56951(2) \\
       	 $y$                 &    0.99282(2)        &      0.99291(3)    &      0.99301(2)  & 0.99314(2) & 0.99314(2) & 0.99316(2) & 0.99317(2) \\
      	 $z$                 &    0.85236(1)        &      0.85232(1)    &      0.85224(1)  & 0.85220(1) & 0.85216(1) & 0.85213(1) & 0.85211(1) \\
				 $U_{\rm eq}$ (\AA$^2$) &   0.00807(5)     &      0.00702(7)   &     0.00604(4) & 0.00498(4) & 0.00443(4) & 0.00396(4) & 0.00330(3) \\ 
    \multicolumn{8}{c}{S2, $4b$: $\frac{3}{4}, \frac{1}{4}, 0$}  \\
				 $U_{\rm eq}$ (\AA$^2$) &   0.00773(8)     &      0.00642(11)   &     0.00586(6) & 0.00490(6) & 0.00443(6) & 0.00391(7) & 0.00335(7) \\ 
       \multicolumn{8}{c}{O1, $16g$: $x, y, z$ }  \\
       	 $x$                 &    0.42699(8)        &      0.42698(11)    &      0.42706(7)  & 0.42711(7) & 0.42705(7) & 0.42715(7) & 0.42713(7) \\
       	 $y$                 &    0.95323(9)        &      0.95341(13)    &      0.95309(8)  & 0.95318(7) & 0.95311(7) & 0.95310(7) & 0.95303(8) \\
      	 $z$                 &    0.83787(5)        &      0.83784(7)    &      0.83763(4)  & 0.83761(4) & 0.83758(4) & 0.83751(4) & 0.83749(4) \\
				 $U_{\rm eq}$ (\AA$^2$) &   0.01278(18)     &      0.01127(27)   &     0.00955(13) & 0.00786(13) & 0.00707(13) & 0.00628(13) & 0.00548(14) \\ 
       \multicolumn{8}{c}{O2, $16g$: $x, y, z$ }  \\
       	 $x$                 &    0.66403(9)        &      0.66447(12)    &      0.66458(7)  & 0.66486(7) & 0.66494(7) & 0.66510(7) & 0.66515(7) \\
       	 $y$                 &    0.92900(8)        &      0.92933(11)    &      0.92940(7)  & 0.92957(7) & 0.92963(6) & 0.92964(7) & 0.92978(7) \\
      	 $z$                 &    0.80658(5)        &      0.80650(7)    &      0.80641(4)  & 0.80630(4) & 0.80623(4) & 0.80619(4) & 0.80616(4) \\
				 $U_{\rm eq}$ (\AA$^2$) &   0.01299(18)     &      0.01094(26)   &     0.00953(13) & 0.00791(13) & 0.00706(13) & 0.00627(13) & 0.00544(14) \\ 
\hline
           \multicolumn{8}{c}{to be continued in Table \ref{Table_XRD2}}  \\
			 \end{tabular}
\end{ruledtabular}
\label{Table_XRD}
\end{table*}

\section{Results and Discussion}

\begin{table*}[t]
\caption{Continued from Table \ref{Table_XRD}.}
\begin{ruledtabular}
		\begin{tabular}[t]{lccccccc}
                $T$ (K)     &             300          &          250       &       200     &   150  &    125  &    100  &    80  \\ \hline
       \multicolumn{8}{c}{O3, $16g$: $x, y, z$ }  \\
       	 $x$                 &    0.58626(9)        &      0.58654(13)    &      0.58651(7)  & 0.58657(7) & 0.58661(7) & 0.58660(7) & 0.58661(8) \\
       	 $y$                 &    0.14226(8)        &      0.14238(11)    &      0.14266(7)  & 0.14282(7) & 0.14293(7) & 0.14294(7) & 0.14302(7) \\
      	 $z$                 &    0.84431(5)        &      0.84413(7)    &      0.84416(4)  & 0.84410(4) & 0.84406(4) & 0.84405(4) & 0.84403(4) \\
				 $U_{\rm eq}$ (\AA$^2$) &   0.01305(18)     &      0.01135(27)   &     0.00951(13) & 0.00784(13) & 0.00709(13) & 0.00634(13) & 0.00539(14) \\ 
       \multicolumn{8}{c}{O4, $16g$: $x, y, z$ }  \\
       	 $x$                 &    0.60434(9)        &      0.60444(12)    &      0.60480(7)  & 0.60492(7) & 0.60498(7) & 0.60502(7) & 0.60511(7) \\
       	 $y$                 &    0.95295(9)        &      0.95292(13)    &      0.95318(8)  & 0.95313(7) & 0.95317(7) & 0.95323(7) & 0.95316(8) \\
      	 $z$                 &    0.92021(5)        &      0.92031(7)    &      0.92030(4)  & 0.92030(4) & 0.92030(4) & 0.92026(4) & 0.92029(4) \\
				 $U_{\rm eq}$ (\AA$^2$) &   0.01303(18)     &      0.01093(26)   &     0.00967(13) & 0.00787(13) & 0.00709(12) & 0.00631(13) & 0.00549(13) \\ 
       \multicolumn{8}{c}{O5, $16g$: $x, y, z$ }  \\
       	 $x$                 &    0.65483(9)        &      0.65489(12)    &      0.65492(7)  & 0.65498(7) & 0.65494(7) & 0.65494(7) & 0.65494(7) \\
       	 $y$                 &    0.33044(9)        &      0.33067(12)    &      0.33088(7)  & 0.33097(7) & 0.33115(7) & 0.33123(7) & 0.33123(7) \\
      	 $z$                 &    0.95956(5)        &      0.95942(7)    &      0.95948(4)  & 0.95943(4) & 0.95941(4) & 0.95941(4) & 0.95942(4) \\
				 $U_{\rm eq}$ (\AA$^2$) &   0.01390(18)     &      0.01233(28)   &     0.01011(13) & 0.00829(14) & 0.00736(13) & 0.00661(13) & 0.00577(14) \\ 
       \multicolumn{8}{c}{O6, $16g$: $x, y, z$ }  \\
       	 $x$                 &    0.36485(7)        &      0.36470(10)    &      0.36460(6)  & 0.36440(6) & 0.36439(6) & 0.36435(6) & 0.36431(7) \\
       	 $y$                 &    0.39040(8)        &      0.39061(10)    &      0.39063(7)  & 0.39082(6) & 0.39089(7) & 0.39100(7) & 0.39105(7) \\
      	 $z$                 &    0.95094(4)        &      0.95092(6)    &      0.95097(4)  & 0.95095(4) & 0.95095(4) & 0.95098(4) & 0.95095(4) \\
				 $U_{\rm eq}$ (\AA$^2$) &   0.00811(14)     &      0.00728(22)   &     0.00630(11) & 0.00544(11) & 0.00498(11) & 0.00462(12) & 0.00411(13) \\ 
    \multicolumn{8}{c}{K, $4a$: $\frac{3}{4}, \frac{1}{4}, \frac{3}{4}$}  \\
				 $U_{\rm eq}$ (\AA$^2$) &   0.05965(26)     &      0.05479(34)   &     0.04851(18) & 0.04224(16) & 0.03872(14) & 0.03539(14) & 0.03230(14) \\ 
    \multicolumn{8}{c}{Cl, $4c$: $\frac{1}{4}, \frac{1}{4}, z$}  \\
      	 $z$                 &    0.82217(3)        &       0.82208(4)    &      0.82221(2)  & 0.82226(2) & 0.82226(2) & 0.82231(2) & 0.82234(3) \\
				 $U_{\rm eq}$ (\AA$^2$) &   0.01266(7)     &      0.01058(11)   &     0.00892(6) & 0.00720(6) & 0.00634(5) & 0.00565(6) & 0.00486(6) \\ \hline
           \multicolumn{8}{c}{Goodness of fit and R values}  \\
			 GOF                       &       1.99           &         2.33        &       1.84 &  1.87  & 1.88 & 1.96 &  2.11 \\
			 $wR_2$ (\%)               &       6.42           &         7.57        &       5.61 & 5.60  & 5.65  & 5.97 & 6.38 \\
			 $R_1$ (\%)                &       2.53           &         3.20        &       2.07  & 2.04 & 2.05 & 2.25 &  2.48 \\ 
			 \end{tabular}
\end{ruledtabular}
\label{Table_XRD2}
\end{table*}

\begin{figure*}
\includegraphics[scale= 0.43]{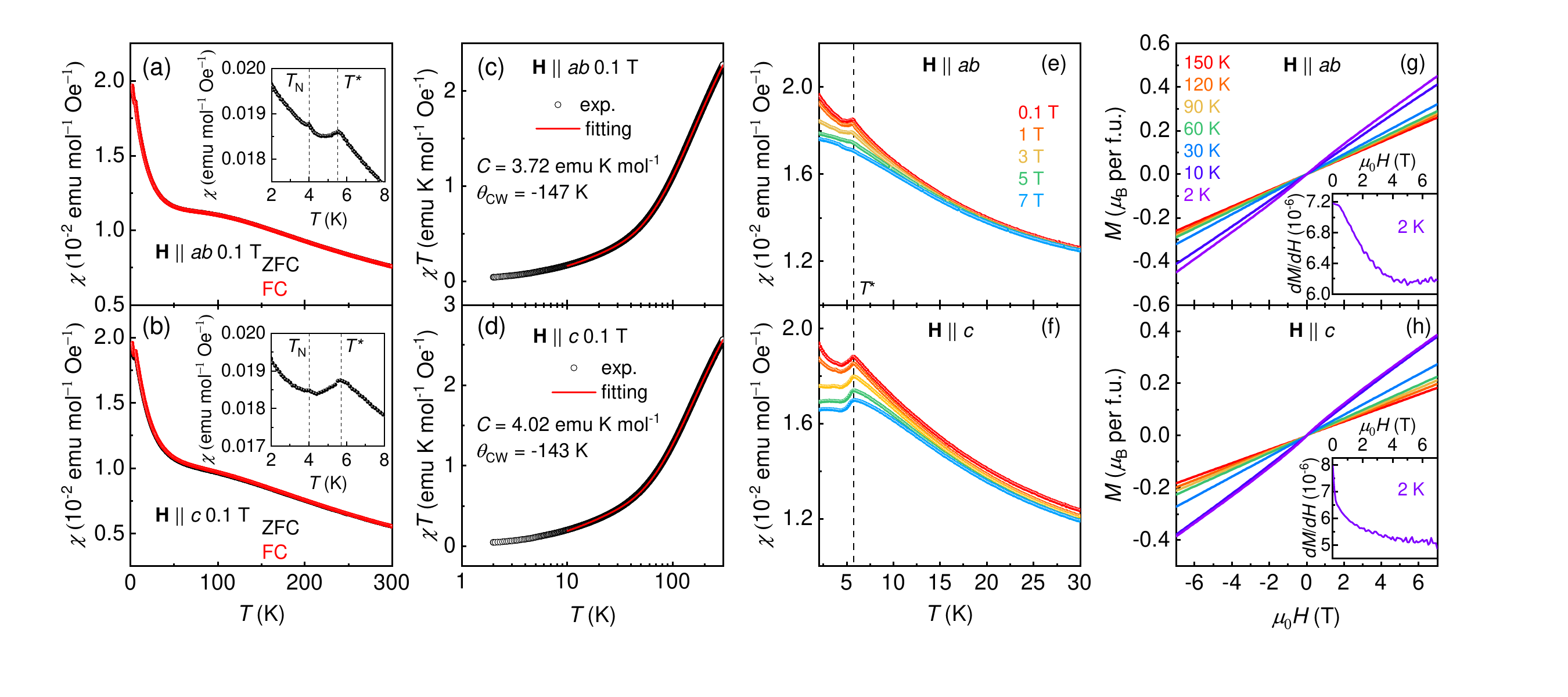}
\caption{ (a) and (b) Temperature dependence of the magnetic susceptibility measured for a KCu$_7$TeO$_4$(SO$_4$)$_5$Cl single crystal with $H\parallel ab$ and $H\parallel c$, respectively. Red and black curves represent field-cooled (FC) and zero-field-cooled (ZFC) measurements. Insets in (a) and (b) show enlarged view at low temperatures near the transitions at $T^*$ and $T_\mathrm{N}$. (c) and (d) Theoretical fitting (red lines) of the magnetic susceptibility using the Husimi ansatz shown in Eq. \ref{eq1}. (e) and (f) Low-temperature magnetic susceptibility measured in various external magnetic fields.  (g) and (h) Isothermal magnetization measured at selected temperatures. Insets in (g) and (h): Derivative of the magnetization obtained at 2 K.  }
\label{fig2}
\end{figure*}

The nabokoite KCu$_7$TeO$_4$(SO$_4$)$_5$Cl crystallizes into a tetragonal structure (space group $P4/ncc$), as shown in Fig.\@ \ref{fig1}(d).\@ When all atoms are considered, the structure of KCu$_7$TeO$_4$(SO$_4$)$_5$Cl is quite complex, consisting of an intricate network formed by different coordination polyhedra between K, Cu, Te, and S, on one hand, and O and Cl, on the other.
However, its physical properties are primarily governed by the network of Cu1 and Cu3 atoms, illustrated from different perspectives in Fig.\@ \ref{fig1}(b) and (c). Although this Cu network is significantly corrugated, it closely resembles the theoretical planar SKL plaquettes composed of Cu1 squares and Cu1-Cu3-Cu1 triangles, as depicted in Fig.\@ \ref{fig1}(a).\@ In the experimental motif, however, the triangles are not isosceles and therefore exhibit different atomic distances of $d_1{\rm (Cu1-Cu1)}$ = 3.275 \AA, $d_2{\rm (Cu1-Cu3)}$ = 3.094 \AA, and $d_3{\rm (Cu1-Cu3)}$ = 3.500 \AA.\@ The diagonal distance between two Cu1 atoms of the squares is $d_9{\rm (Cu1-Cu1)}$ = 4.631 \AA.\@ In addition, the Cu1 squares are rotated around the $c$ axis by an angle of 5.90$^{\circ}$ with respect to the $a$ direction of the tetragonal unit cell. Above and below the centers of the Cu1 squares within the distorted plaquettes, decorating Cu2 atoms are situated, whose distances to the corresponding Cu1 and Cu3 atoms are $d_4{\rm (Cu2-Cu1)}$  = 5.371 {\AA} and $d_5{\rm (Cu2-Cu3)}$ = 6.677 \AA,\@ on the side from which the triangles are oriented away, and $d_6{\rm (Cu2-Cu1)}$ =  7.063 {\AA} and $d_7{\rm (Cu2-Cu3)}$ = 4.619 {\AA} on the side where the triangles are oriented towards the Cu2 atoms. Due to the four-fold axis passing through the Cu2 atoms, their interactions with the atoms in the plaquettes are expected to be essentially frustrated, in close analogy to the frustration observed among the Cu1 and Cu3 atoms within the triangular and square motifs of the distorted plaquettes.

Based on the crystal field arising from the distortion of the Cu1O$_5$Cl$_1$ and Cu3O$_6$ octahedra, the Cu $3d_{x^2-y^2}$ states are the key orbitals occupied by electrons at the Fermi level. Therefore, their superexchange interactions, mediated via Cu–O–Cu bonding angles, govern the nearest-neighbor magnetism. The angles determined from the structural data are $\varphi_1{\rm (Cu1-O6-Cu1)} \approx 109.30^\circ$, $\varphi_2{\rm (Cu1-O6-Cu3)} \approx 101.95^\circ$, and $\varphi_3{\rm (Cu1-O6-Cu3)} \approx 119.63^\circ$,\@ where the largest (smallest) angle corresponds to the strongest (weakest) antiferromagnetic interaction.\@ Here, the largest (smallest) bond angle also corresponds to the longest (shortest) distance in the non-isosceles Cu1-Cu3-Cu1 triangle shown in Fig.\@ \ref{fig1}(b) and (c).\@ Upon cooling from 300 to 80 K,\@ all lattice parameters and the volume, together with the anisotropic displacement parameters of the atoms decrease continuously and for example, the above-mentioned Cu–Cu distances change at 80 K slightly to 
$d_1{\rm (Cu1-Cu1)}$ = 3.267 \AA, $d_2{\rm (Cu1-Cu3)}$ = 3.083 \AA, $d_3{\rm (Cu1-Cu3)}$ = 3.449 \AA,\@ and $d_9{\rm (Cu1-Cu1)}$ = 4.621 \AA, and the corresponding bond angles to $\varphi_1{\rm (Cu1-O6-Cu1)} \approx 109.27^\circ$, $\varphi_2{\rm (Cu1-O6-Cu3)} \approx 101.65^\circ$, and $\varphi_3{\rm (Cu1-O6-Cu3)} \approx 120.06^\circ$.\@ 
Furthermore, at 80 K, the distances between the decorating Cu2 atoms and the corresponding Cu1 and Cu3 atoms exhibit only small temperature-dependent \@ variations: on the side where the triangles are oriented away, the distances are $d_4{\rm (Cu2-Cu1)}$ = 5.364 \AA\ and $d_5{\rm (Cu2-Cu3)}$ = 6.668 \AA, respectively whereas on the side where the triangles are oriented towards Cu2, the distances are $d_6{\rm (Cu2-Cu1)}$ = 7.037 \AA\ and $d_7{\rm (Cu2-Cu3)}$ = 4.597 \AA.

The Cu2 atom is in a pyramidal Cu2O$_4$Cl$_1$ coordination, with four basal O atoms and one apical Cl atom. The latter is shared by the Cu2 atom with the two adjacent Cu1O$_5$Cl$_1$ octahedra, forming an angle of $\varphi_{\rm Cl}{\rm (Cu1-Cl-Cu2)} \approx 122.77^\circ$,\@ in principle, allowing for a magnetic superexchange interaction between Cu1 and Cu2.\@ Interestingly, the Cu2 atoms are also connected to the next neighboring plaquettes which are positioned somewhat laterally to them in Fig.\@ \ref{fig1}(c) via SO$_4$ tetrahedra [see also Fig.\@ \ref{fig1}(d)],\@ with a relatively short distance of $d_8{\rm (Cu2-Cu1)} = 5.679$ \AA. Upon cooling to 80 K, $\varphi_{\rm Cl}{\rm (Cu1-Cl-Cu2)}$ and $d_8{\rm (Cu2-Cu1)}$ are changed to 122.65$^\circ$ and 5.664 \AA,\@ respectively. In this sense, a quasi 3D network is formed structurally, potentially also facilitating interactions between the corrugated plaquettes, mediated by the Cu2 atoms. For specific temperatures, the relevant structural parameters, along with atomic positions and thermal displacement parameters, are summarized in Tables \ref{Table_XRD} and \ref{Table_XRD2}.\@ Using \textsc{Pascal} \cite{cliffe2012pascal}, thermal expansion coefficients were determined from temperature-dependent measurements between 300 and 80 K:  The coefficients were found to be \( 8.4(1) \times 10^{-6} \, \text{K}^{-1} \) for the \( a \) axis, \( 13.7(5) \times 10^{-6} \, \text{K}^{-1} \) for the \( c \) axis, and \( 30.5(7) \times 10^{-6} \, \text{K}^{-1} \) for the volumetric expansion.

Figure~\ref{fig2} presents the temperature-dependent magnetic susceptibility ($\chi$) of a KCu$_7$TeO$_4$(SO$_4$)$_5$Cl single crystal. For both in-plane ($H\parallel ab$) and out-of-plane ($H\parallel c$) field configurations, $\chi(T)$ exhibits similar behavior with slight anisotropy. No significant difference between zero-field-cooled (ZFC) and field-cooled (FC) measurements is observed below 300 K. In the high-temperature paramagnetic regime, a broad hump appears around 100 K, consistent with previous reports on polycrystalline samples \cite{markina2024static,murtazoev2023new}. This behavior deviates markedly from the standard Curie-Weiss law and is commonly observed in frustrated magnets, where frustration leads to a breakdown of conventional mean-field descriptions \cite{PohleCW}. To better describe this deviation, Pohle and Jaubert proposed a generalized Husimi ansatz \cite{PohleCW}:
\begin{equation}
\chi T = \frac{1 + b_1 \mathrm{exp}(c_1/T)}{a + b_2 \mathrm{exp}(c_2/T)},
\label{eq1}
\end{equation}
from which the Curie-Weiss temperature ($\theta_\mathrm{CW}$) and Curie constant ($C$) can be extracted as:
\begin{equation}
\theta_\mathrm{CW} = \frac{b_1 c_1}{1 + b_1} - \frac{b_2 c_2}{a + b_2}, \quad C = \frac{1 + b_1}{a + b_2}.
\label{eq2}
\end{equation} As shown in Figs.~\ref{fig2}(c) and \ref{fig2}(d), this ansatz provides an excellent fit to the reduced susceptibility $\chi T$ over a wide temperature range (10–300 K). For $H\parallel ab$ ($H\parallel c$), the extracted values are $\theta^a_\mathrm{CW} = -147$ K ($\theta^c_\mathrm{CW}=-143$ K) and $C = 3.72$ emu·K·mol$^{-1}$ ($4.02$ emu·K·mol$^{-1}$). Comparable values, $\theta_\mathrm{CW} = -153.6$ K and $C = 3.84$ emu·K·mol$^{-1}$, have also been reported for powder samples \cite{Gonzalez2025}. Assuming spin-only contributions ($S=1/2$ for Cu$^{2+}$ ions), the $g$-factor can be estimated from $C = ng^2 S(S + 1)/8$ with $n = 7$ Cu$^{2+}$ ions per unit cell, yielding nearly isotropic values: $g = 2.38$ for $H\parallel ab$ and $g = 2.47$ for $H\parallel c$. Electron spin resonance measurements on powder samples revealed similar slightly anisotropic $g$-factors of 2.07 and 2.33 at 20 K \cite{markina2024static}, which likely reflect an average over the anisotropic ligand environments of Cu$^{2+}$ ions in Cu1O$_5$Cl$_1$, Cu3O$_6$ octahedra, and Cu2O$_4$Cl$_1$ pyramids. We note that the Curie-Weiss temperatures along the $c$ and $a$ axes are related to the exchange interactions through \cite{WANG1971,Boutron1973}:
\begin{equation}
  \theta^c_\mathrm{CW}=-\frac{J(J+1)}{3k_B}J^c_\mathrm{ex}-\frac{(2J-1)(2J+3)}{5k_B}B^0_2,
\end{equation}
\begin{equation}
    \theta^a_\mathrm{CW}=-\frac{J(J+1)}{3k_B}J^a_\mathrm{ex}+\frac{(2J-1)(2J+3)}{10k_B}B^0_2,
\end{equation}
where $J$ is the total angular momentum of the Cu$^{2+}$ ion, $J^c_\mathrm{ex}$ and $J^a_\mathrm{ex}$ denote the sums of exchange interactions along the $c$ and $a$ axes, respectively, and $B^0_2$ is a crystal-field-related parameter. Positive values of $J^c_\mathrm{ex}$ and $J^a_\mathrm{ex}$ correspond to antiferromagnetic interactions. Although $B^0_2$ (typically on the order of 1 K) is unknown in the present case, its contribution to $\theta_\mathrm{CW}$ is negligible assuming spin only contribution with $J=S=1/2$. The nearly isotropic Curie–Weiss temperatures therefore indicate that the summed exchange interactions along the $c$ and $a$ directions are comparable. This is consistent with the theoretical results of M. G. Gonzalez \textit{et al.} \cite{Gonzalez2025}, who showed that exchange couplings with substantial out-of-plane components—such as Cu1–Cu3 [$\sim165$ K, with bond distance labeled as $d_2$ in Fig.~\ref{fig1}(b)] and Cu1–Cu2 [$\sim152$ K, bond distance $d_4$ displayed in Fig.~\ref{fig1}(c)]—are comparable in magnitude to the dominant in-plane Cu1–Cu1 interactions [$\sim118$ K and $\sim170$ K, with bond distances $d_1$ and $d_9$ shown in Fig.~\ref{fig1}(b)]. Furthermore, although the interlayer Cu2–Cu3 exchange coupling [$\sim -9$ K, bond distance $d_7$ shown in Fig.~\ref{fig1}(c)] amounts to only about 5\% of the largest intralayer Cu1–Cu1 interaction ($\sim170$ K along the diagonal $d_9$ bond of the Cu1 squares), neglecting this term leads to a qualitatively different magnetic ground state \cite{Gonzalez2025}. Only by including the interlayer interactions can a transition temperature consistent with experimental observations in zero magnetic field be reproduced. These results highlight the important roles played by the decorating Cu2 sites in determining the exchange constants and magnetic properties of this system, and demonstrate that a purely two-dimensional model is insufficient to fully capture the magnetic behavior of KCu$_7$TeO$_4$(SO$_4$)$_5$Cl.

At low temperatures, a cusp-like anomaly emerges around $T^*\sim5.6$ K [see insets in Figs.~\ref{fig2}(a) and \ref{fig2}(b)], reminiscent of the antiferromagnetic (AFM) transition reported in polycrystalline samples \cite{markina2024static,murtazoev2023new}. However, $T^*$ significantly exceeds the AFM transition temperature $T_\mathrm{N}\sim3.2$ K observed in powder samples. Instead, $T^*$ closely matches the temperature range ($\sim$5.5–5.7 K) where an unidentified specific heat anomaly appears in polycrystals \cite{markina2024static,murtazoev2023new}, suggesting that the anomaly at $T^*$ does not correspond to the AFM transition identified in those studies. Notably, a second, weaker kink is visible in the magnetic susceptibility near $T_\mathrm{N} \sim 4$ K, likely marking the AFM transition. This interpretation is further supported by the NMR results discussed below. The large frustration parameter, $f = |\theta_\mathrm{CW}|/T_\mathrm{N} > 35$, underscores the potential of this system as a platform for exploring strongly frustrated magnetism. Above $T_\mathrm{N}$ in the paramagnetic regime, the magnetization $M$ varies linearly with magnetic field for both field directions [Figs.~\ref{fig2}(g,h)]. Below $T_\mathrm{N}$ at 2 K, the $M$–$H$ curves show only minor deviations from linearity and no hysteresis upon increasing or decreasing field. In contrast to previous observations in powder samples, where a sizable residual magnetization ($\sim$0.3 $\mu_\mathrm{B}$ per formula unit) was attributed to $\sim$4\% impurity contributions \cite{markina2024static}, our single-crystal data show negligible remanent magnetization at zero field, indicating the high sample quality. Furthermore, two field-induced metamagnetic transitions were previously observed in polycrystals below $T_\mathrm{N}$ \cite{markina2024static}. However, in our single-crystal samples, the differential magnetization $dM/dH$ remains smooth and featureless up to 7 T for both field orientations [insets of Figs.~\ref{fig2}(g,h)], indicating the absence of metamagnetic transitions. These findings suggest that the magnetic ground state in single crystals differs from that in polycrystalline samples. The field-induced metamagnetic transitions observed in powder samples are unlikely to originate from paramagnetic impurities, which would be expected to respond smoothly and monotonically to an applied magnetic field. Orientational averaging effects intrinsic to powder measurements can also be ruled out, as similar field-induced transitions are absent in single crystals for magnetic fields applied both in-plane and out-of-plane. Given the complex exchange network formed by Cu1 and Cu3 ions, together with the decorating Cu2 sites, the magnetic interactions are expected to be highly sensitive to subtle variations in bond lengths and bond angles. We therefore attribute the contrasting field-induced responses observed in powder and single-crystal samples primarily to strain effects in the powder specimens. Such strain may arise either intrinsically from defects such as vacancies or disorder, or extrinsically from mechanical stress introduced during the preparation of compressed powder pellets for measurements. This interpretation is further supported by the slightly different magnetic transition temperatures observed in powder ($T_\mathrm{N} \sim 3.2$ K) and single-crystal samples ($T_\mathrm{N} \sim 4.5$ K; see below). Taken together, these results indicate that strain or pressure constitutes a prominent tuning parameter for tailoring the magnetic properties of KCu$_7$TeO$_4$(SO$_4$)$_5$Cl.

\begin{figure}[t]
\centering
\includegraphics[width=8.5cm]{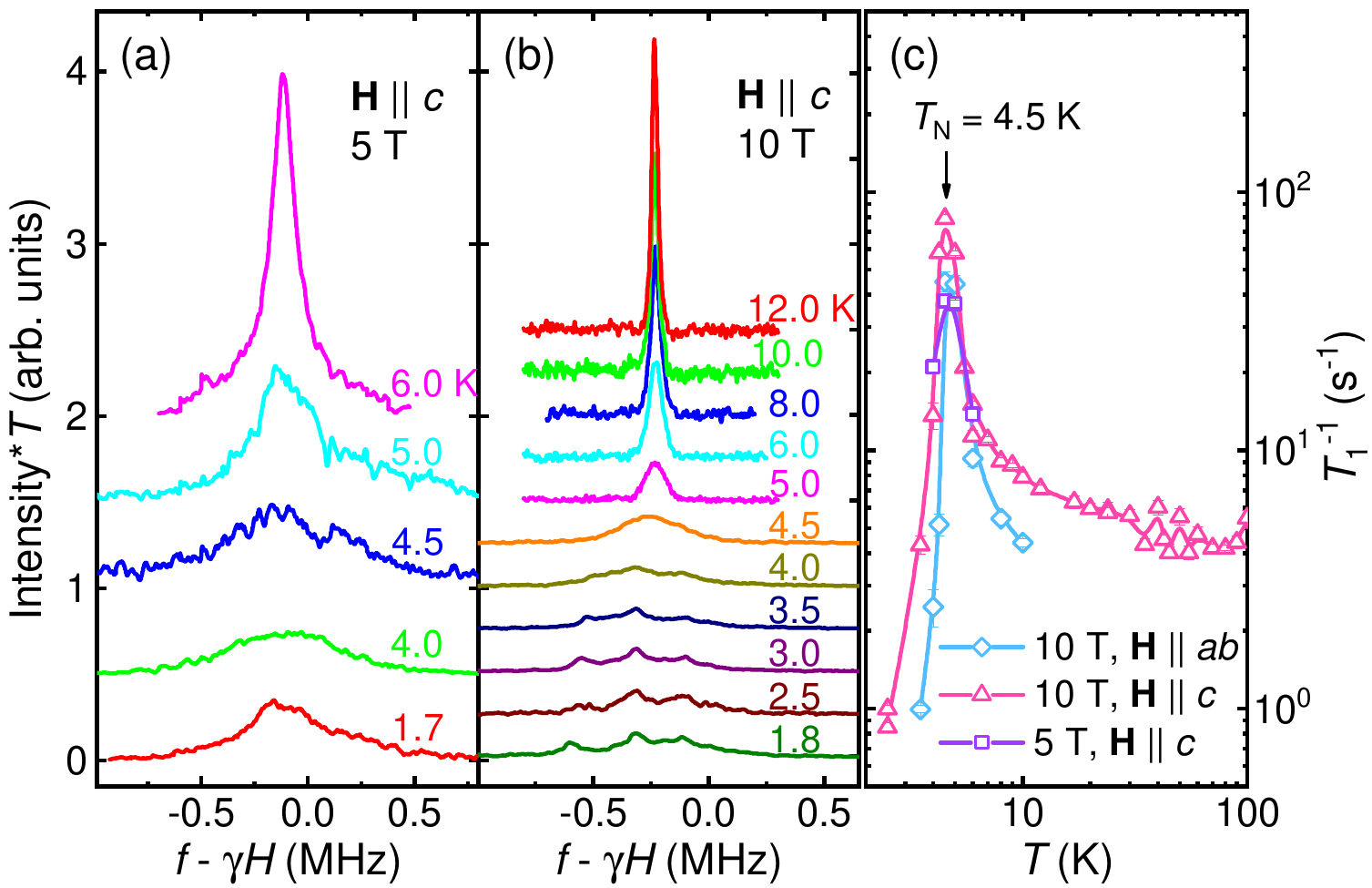}
\caption{\label{nmr}
NMR spectra and spin-lattice relaxation rate $1/T_1$.
(a,b) $^{35}$Cl spectra measured at various temperatures with fields of 5~T and 10~T applied along the $c$ axis. Data are shifted vertically for clarity.
(c) $1/T_1$ as functions of temperatures measured under fields of 5~T and 10~T under two field orientations as labeled.
}
\end{figure}

Our measured NMR spectra and the spin-lattice relaxation rates provide unambiguous spectroscopic evidence for the emergence of the AFM order at low temperatures. Figure~\ref{nmr}(a)-(b) shows the $^{35}$Cl NMR spectra collected at various temperatures under magnetic fields of 5~T and 10~T, applied along the crystalline $c$ axis. At temperatures at and above 6~K, the spectra display a single sharp peak, indicating a uniform paramagnetic phase at high temperatures. Upon cooling through 5~K, the resonance lines broaden significantly in both fields, reflecting the development of strong magnetic fluctuations. Notably, under 10~T field, the NMR line split into three peaks when cooled below 4~K, with frequencies shift to both negative and positive directions. Such a line split is a direct evidence of the onset of static AFM ordering, where both negative and positive hyperfine field is identified on the $^{35}$Cl site.

To further verify the magnetic phase transitions, the spin-lattice relaxation rate $1/T_1$ is measured at 5~T and 10~T,  with field in the $ab$ plane and along the $c$ axis, and the data are shown as functions of temperatures in Fig.~\ref{nmr}(c). For $H \parallel c$ at 10~T, $1/T_1$ stays nearly constant at temperatures above 30~K, consistent with paramagnetic behavior. Upon cooling, $1/T_1$ increases gradually, marking the onset of strong low-energy spin fluctuations. A pronounced peak is formed at 4.5~K, which characterizes the transition  temperature $T_{\rm N}$ of magnetic ordering. Similar peaked behaviors are also observed for the out-of-plane field 5~T and the in-plane field 10~T. However, $T_N$ stays unchanged at 4.5~K under these fields, which indicates the magnetic coupling has an effective energy scale much higher than 10~T, in consistency with the large Curie-Weiss temperature observed above.

\begin{figure}[t]
\includegraphics[scale= 0.4]{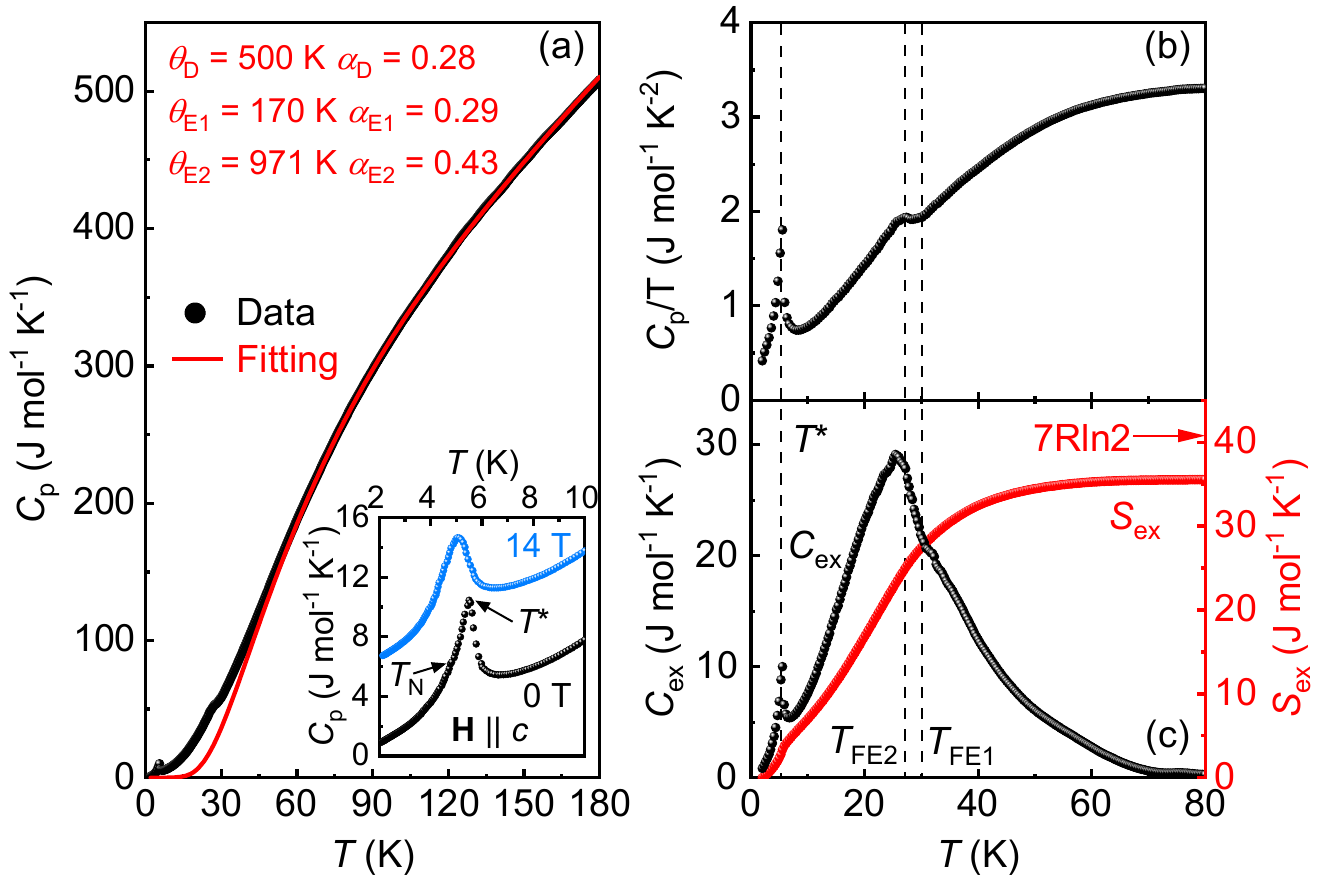}
\caption{ (a) Specific heat of a KCu$_7$TeO$_4$(SO$_4$)$_5$Cl single crystal. The red line is a theoretical fitting to the phonon specific heat using the Debye Einstein model shown in Eq. \ref{eq3}. Inset in (a): Low-temperature specific heat near the transitions at $T^*$ and $T_\mathrm{N}$ in 0 and 14 T ($H\parallel c$). The curve for 14 T has been shift vertically for clarity. (b) Low-temperature specific heat divided by temperature $C_\mathrm{p}/T$, showing clear transitions at $T_\mathrm{FE1}$, $T_\mathrm{FE2}$ and $T^*$. (c) Excess specific heat ($C_\mathrm{ex}$) and the corresponding entropy ($S_\mathrm{ex}$). The excess specific heat is obtained by subtracting the total specific heat from the estimated phonon background shown in (a).   }  
\label{fig3}
\end{figure}

\begin{figure*}[t]
\includegraphics[scale= 0.525]{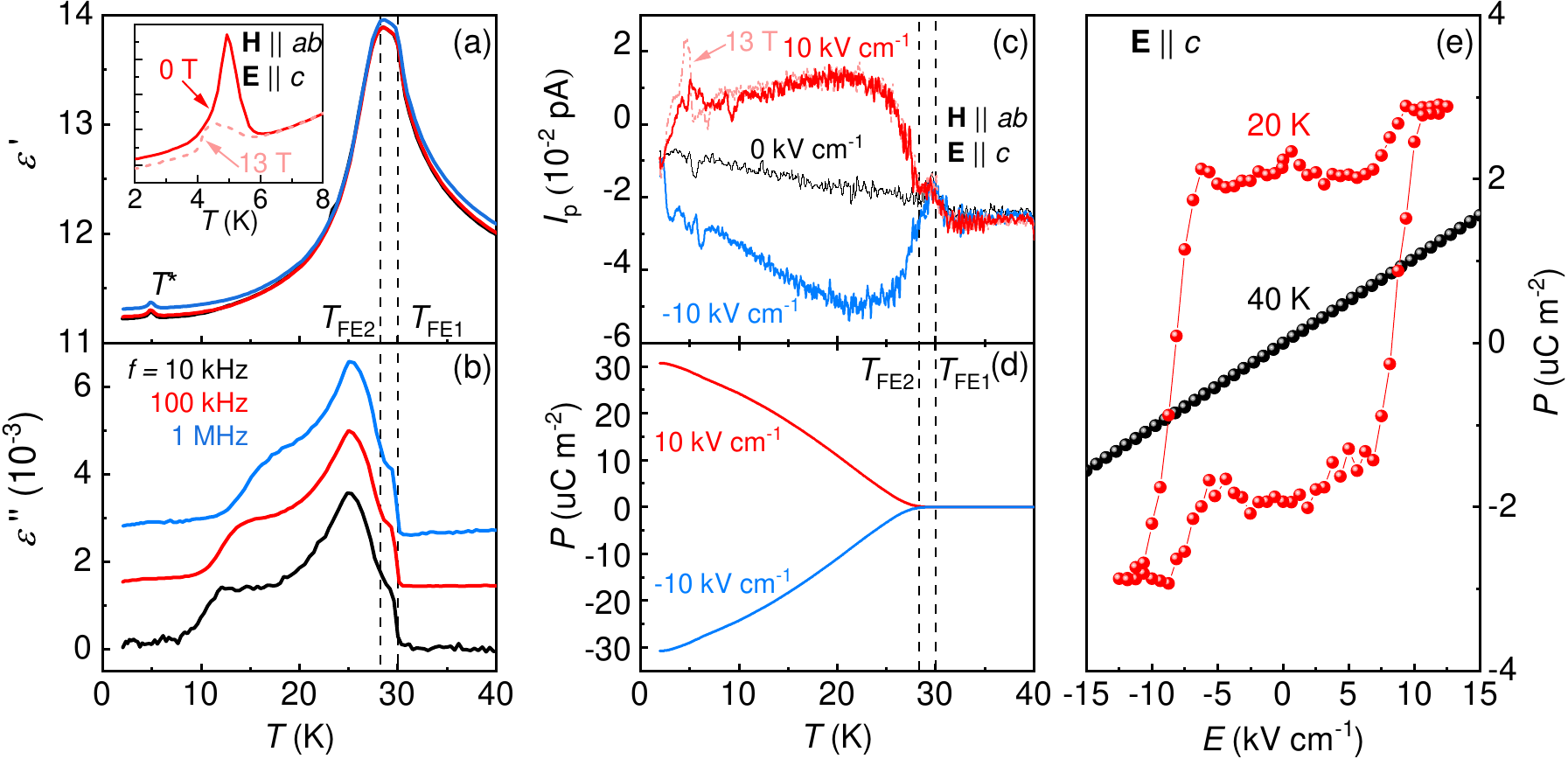}
\caption{ (a) and (b) Temperature dependence of the real ($\varepsilon'$) and imaginary ($\varepsilon''$) parts of the relative dielectric constant measured using three frequencies. (c) and (d) Temperature-dependent pyroelectric current ($I_\mathrm{p}$) and electric polarization ($P$) measured using opposite electric fields ($E\parallel c=\pm10$ kV cm$^{-1}$). Background current measured in zero electric field is also presented for comparison. (e) $P-E$ hysteresis loops measured at 20 and 40 K. }  
\label{fig4}
\end{figure*}

Figure~\ref{fig3}(a) shows the specific heat ($C_\mathrm{p}$) of a KCu$_7$TeO$_4$(SO$_4$)$_5$Cl single crystal. At high temperatures, $C_\mathrm{p}$ is well described by a phonon model consisting of one Debye and two Einstein terms [red line in Fig.~\ref{fig3}(a)]:

\begin{multline}
\begin{split}
C_\mathrm{ph}(T) = &9N\alpha_\mathrm{D} R (\frac{T}{\Theta_\mathrm{D}})^3\int_{0}^{\Theta_\mathrm{D}/T}\frac{x^{4}e^{x}}{(e^{x}-1)^{2}}dx \\
&+3N\alpha_\mathrm{E1} R(\frac{\Theta_\mathrm{E1}}{T})^{2}\frac{e^{\Theta_\mathrm{E1}/T}}{(e^{\Theta_\mathrm{E1}/T}-1)^{2}}\\&+3N\alpha_\mathrm{E2} R(\frac{\Theta_\mathrm{E2}}{T})^{2}\frac{e^{\Theta_\mathrm{E2}/T}}{(e^{\Theta_\mathrm{E2}/T}-1)^{2}}.
\label{eq3}
\end{split}
\end{multline}

Here, $x=\hbar\omega/k_B T$, $R$ is the gas constant, and $N = 39$ is the number of atoms per formula unit. The best-fit parameters are: $\Theta_\mathrm{D}=500$ K, $\Theta_\mathrm{E1}=170$ K, $\Theta_\mathrm{E2}=971$ K, with weighting factors $\alpha_\mathrm{D}=0.28$, $\alpha_\mathrm{E1}=0.29$, and $\alpha_\mathrm{E2}=0.43$ (summing to 1). These values are consistent with those for powder samples \cite{markina2024static}. At low temperatures, deviations from the model occur due to multiple transitions. A sharp peak is seen at $T^* = 5.5$ K, while only a subtle kink appears near $T_\mathrm{N} \sim 4.5$ K [see inset of Fig.~\ref{fig3}(a)]. The $T^*$ peak is weakly field-dependent; it slightly broadens and shifts by only 0.5 K in a 14 T field ($H \parallel c$). A similar anomaly at $T^* \sim 5.5$–5.7 K is also reported in powder samples \cite{markina2024static,murtazoev2023new}. The origin of this transition remains unclear. In a spin-1/2 Heisenberg antiferromagnet on an ideal square-kagome lattice (SKL) with $J_1 = J_2 = J$, numerous singlet states exist below the lowest triplet excitations \cite{richter2022thermodynamics,Astrakhantsev_singlet}. Their excitations can produce a shoulder or broad peak in $C_\mathrm{p}$ at $T \sim 0.05J$–$0.2J$ \cite{richter2022thermodynamics,Tomczak_2003}. However, magnetic fields are expected to sharpen and shift this peak, in contrast to the experimental observations. Vacancy-induced effects in frustrated magnets can also give rise to broad $C_\mathrm{p}$ anomalies \cite{Sedik2025_vacancy}. However, since both single crystals and the entire $A$Cu$_7$TeO$_4$(SO$_4$)$_5$Cl powder series exhibit similar features at $T^*$, this would require nearly identical defect concentrations across all samples—a scenario unlikely given differences in growth conditions.

In addition to the transitions at $T_\mathrm{N}$ and $T^*$, two additional anomalies are observed at $T_\mathrm{FE1} \sim 30$ K and $T_\mathrm{FE2} \sim 27$ K [see Fig.\ref{fig3}(b)]. In contrast, only a single transition near 25 K, attributed to an antiferroelectric transition, was reported in powder samples \cite{markina2024static}. Therefore, the transitions at $T_\mathrm{FE1}$ and $T_\mathrm{FE2}$ observed in single crystals are likely related to the ordering of electric dipoles. Further details regarding their nature will be discussed in the context of dielectric and pyroelectric measurements [see Fig. \ref{fig4}]. To isolate the contributions from these transitions, we subtracted the estimated phonon background $C_\mathrm{ph}$ from the total specific heat $C_\mathrm{p}$ to obtain the excess specific heat, $C_\mathrm{ex} = C_\mathrm{p} - C_\mathrm{ph}$, as shown in Fig.~\ref{fig3}(c). $C_\mathrm{ex}$ exhibits a peak at $T_\mathrm{FE2}$ and a shoulder at $T_\mathrm{FE1}$, whereas the low-temperature anomalies at $T^*$ and $T_\mathrm{N}$ contribute only weakly. Notably, the excess specific heat persists up to approximately 70 K, suggesting the presence of short-range fluctuations of electric polarization well above the long-range ordering temperatures. The associated excess entropy, estimated by integrating $C_\mathrm{ex}/T$, is plotted as the red line in Fig.~\ref{fig3}(c). Evidently, the transitions at $T_\mathrm{FE1}$ and $T_\mathrm{FE2}$ account for the majority of the entropy release, while those at $T^*$ and $T_\mathrm{N}$ contribute only marginally. The total magnetic entropy is expected to be $S_\mathrm{mag} = 7R\mathrm{ln}2 = 40.3$ J mol$^{-1}$ K$^{-1}$ for seven Cu$^{2+}$ ($S = 1/2$) ions per formula unit, where $R = 8.314$ J mol$^{-1}$ K$^{-1}$ is the gas constant. However, this value is not fully recovered up to 80 K, even when the entropy release at $T_\mathrm{FE1}$ and $T_\mathrm{FE2}$ is included. Considering the large exchange energy scale ($\sim150$ K), significant short-range spin correlations are likely to persist to temperatures well above 100 K. A more accurate estimation of the phonon contribution, for example using a nonmagnetic isostructural analog, would be valuable for quantitatively determining the temperature range over which such spin correlations remain relevant.

To further investigate the transitions at $T_\mathrm{FE1}$ and $T_\mathrm{FE2}$, we present dielectric and pyroelectric current measurements in Fig. \ref{fig4}. As shown in Fig. \ref{fig4}(a), the real part of the dielectric constant ($\varepsilon'$) increases rapidly upon cooling toward $T_\mathrm{FE1}$. Below $T_\mathrm{FE1}$, the slope of $\varepsilon'$ changes slightly and reaches a maximum at $T_\mathrm{FE2}$, followed by a rapid drop upon further cooling. Additionally, a peak in $\varepsilon'$ is observed at $T^*$. All three anomalies at $T_\mathrm{FE1}$, $T_\mathrm{FE2}$, and $T^*$ are frequency-independent up to 1 MHz, indicating true thermodynamic transitions. The imaginary part of the dielectric constant ($\varepsilon''$), or dielectric loss, is nearly constant above $T_\mathrm{FE1}$ [Fig.~\ref{fig4}(b)]. At $T_\mathrm{FE1}$, $\varepsilon''$ increases sharply, develops a shoulder at $T_\mathrm{FE1}$, and continues rising to a peak around 25 K. A plateau-like region emerges between 10 and 20 K, below which $\varepsilon''$ becomes nearly temperature-independent again. These features suggest the emergence of ferroelectric (FE) or antiferroelectric (AFE) order below $T_\mathrm{FE1}$ and $T_\mathrm{FE2}$. To further clarify the nature of these transitions, we performed pyroelectric current measurements [Figs.~\ref{fig4}(c) and \ref{fig4}(d)]. Above $T_\mathrm{FE1}$, the pyroelectric currents measured in opposite electric fields coincide with the background current measured in zero electric field. Just below $T_\mathrm{FE1}$, the currents split and deviate from the background, signaling the onset of spontaneous electric polarization. Below $T_\mathrm{FE2}$, the currents increase rapidly, peaking broadly near 25 K, and return to background levels below 5 K. The sign reversal of the pyroelectric current with opposite electric fields confirms the FE ordering below $T_\mathrm{FE1}$ and $T_\mathrm{FE2}$. The concurrent rise in dielectric loss $\varepsilon''$ below $T_\mathrm{FE1}$ and $T_\mathrm{FE2}$ likely reflects enhanced motion of ferroelectric domain walls. Electric polarization $P$ was obtained by integrating the pyroelectric current after subtracting the background current measured in zero electric field. As shown in Fig.~\ref{fig4}(d), $P$ starts to develop below $T_\mathrm{FE1}$ and grows significantly below $T_\mathrm{FE2}$, with its sign reversible under opposite electric fields, further evidence for FE ordering. The FE state can be probed more directly from the $P-E$ hysteresis loop measurements. As presented in Fig.~\ref{fig4}(e), the electric polarization varies linearly with applied electric field in the paraelectric phase at 40 K. In contrast, a clear $P-E$ hysteresis loop is observed in the FE state at 20 K. 

\begin{figure}[t]
\includegraphics[scale= 0.7]{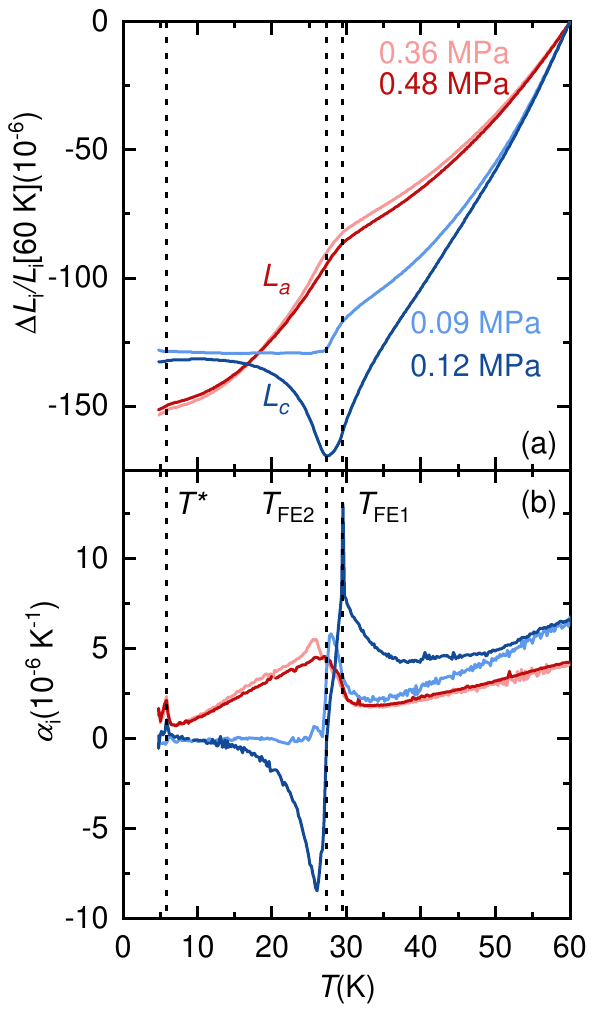}
\caption{(a) Temperature dependence of the linear thermal expansion, $\Delta L_i/L_i$ (referenced to 60 K), measured along the $i = a$ and $c$ axes. Moderate uniaxial pressures were applied by the capacitance dilatometer, which are generally sufficient to prevent structural twinning. Different uniaxial pressures were intentionally employed to highlight the contrasting responses of the $a$ and $c$ axes. (b) Corresponding linear thermal-expansion coefficients, $\alpha_i = (1/L_i)(dL_i/dT)$.}

\label{fig5}
\end{figure}

The FE ordering is necessarily associated with inversion symmetry breaking. Previous electron spin resonance results of powder samples suggest that the tetragonal symmetry is preserved at low temperatures \cite{markina2024static}. The inversion symmetry breaking is thus likely occuring along the $c$-axis. This scenario is further supported by the linear thermal-expansion measurements ($\Delta L_i/L_i$, $i = a, c$) shown in Fig.~\ref{fig5}. The moderate uniaxial pressure applied by the dilatometer is generally sufficient to prevent structural twinning associated with inversion-symmetry breaking. The in-plane relative length change exhibits only minor anomalies at $T_\mathrm{FE1}$ and $T_\mathrm{FE2}$ and shows weak dependence on increasing uniaxial pressure. In contrast, the $c$-axis thermal expansion displays pronounced anomalies near these transitions, with a strong dependence on the applied $c$-axis pressure during the measurements. In particular, under a uniaxial pressure of 0.12 MPa, significant lattice softening develops along the $c$ axis below 50 K and appears to diverge upon approaching the ferroelectric transitions from above. Below the transitions, the $c$-axis lattice hardens again with further cooling. These features provide clear evidence of lattice instabilities occurring predominantly along the $c$ axis. The contrasting behaviors between the $a$ and $c$ directions are also evidently observed in the linear thermal-expansion coefficients, $\alpha_i(T) = (1/L_i)(dL_i/dT)$ [see Fig.~\ref{fig5}(b)]. Given the large interlayer spacing, these observations strongly suggest that the inversion-symmetry breaking is associated with distortions of the  Cu2O$_4$Cl$_1$ pyramids and the adjacent SO$_4$ tetrahedra. The successive FE transitions at $T_\mathrm{FE1}$ and $T_\mathrm{FE2}$ indicate stepwise structural distortions, the microscopic origin of which warrants further investigation. Finally, a broad peak also appears at $T^*$ in the linear thermal expansion coefficient. Note that the dielectric constant also shows a peak at $T^*$ and pyroelectric current drops below  $T^*$, suggesting that this transition also involves electric polarization. Furthermore, compared to the zero-field results, the dielectric peak at $T^*$ becomes noticeably broadened, while the pyroelectric current around $T^*$ is enhanced in 13 T. These findings indicate a sizable magnetoelectric coupling between the spin and charge degrees of freedom. Thus, both spin and charge channels are likely involved in the formation of the broad peak in specific heat, thermal expansion and dielectric constant at $T^*$.    

\section{Conclusions}
In conclusion, we have successfully grown single crystals of the decorated square-kagome compound KCu$_7$TeO$_4$(SO$_4$)$_5$Cl. The antiferromagnetic order below $T_\mathrm{N}\sim4.5$ K is unambiguously revealed by the magnetic susceptibility and the NMR measurements. The Curie-Weiss temperature and Curie constant are consistent with those reported in polycrystals. However, unlike polycrystalline samples, no field-induced metamagnetic transition is observed up to 7~T in the current study. The nearly isotropic magnetization of single crystals for both in-plane and out-of-plane fields highlights the significant roles played the decorating Cu2 sites. Moreover, we found ferroelectric ordering emerging below $T_\mathrm{FE1}\sim30$ K and $T_\mathrm{FE2}\sim27$ K, likely driven by an inversion-symmetry breaking induced by distortions in the Cu2O$_4$Cl$_1$ pyramids and the adjacent SO$_4$ tetrahedra. These results demonstrate that the decorating Cu2 sites play a crucial role in governing both the magnetic and electric properties of KCu$_7$TeO$_4$(SO$_4$)$_5$Cl. \\

\section{acknowledgments}
 This work is supported by National Natural Science Foundation of China (Grant No.~12474141 and No.~12134020), Natural Science Foundation of Chongqing, China (Grant No. CSTB2025NSCQ-GPX0729), Fundamental Research Funds for the Central Universities, China (2025CDJ-IAISYB-034), the Venture and Innovation Support Program for Chongqing Overseas Returnees (Grant No. cx2024007), and Chinesisch-Deutsche Mobilitätsprogamm of Chinesisch-Deutsche Zentrum für Wissenschaftsförderung (Grant No. M-0496). Y.C. acknowledges the support by the National Natural Science Foundation of China (Grant No. 12374081), the Open Research Fund of the Pulsed High Magnetic Field Facility (Grant No. WHMFC2024007), and Huazhong University of Science and Technology.  We would like to thank Guiwen Wang and Yan Liu at the Analytical and Testing Center of Chongqing University and Siegmar Roth and Andre Beck at the Institute for Quantum Materials and Technologies,
Karlsruhe Institute of Technology for their technical assistance.

\end{document}